# Learning a Factorized Orthogonal Latent Space using Encoder-only Architecture for Fault Detection; An Alarm Management Perspective

**Vahid MohammadZadeh Eivaghi**

Department of Electrical Engineering K. N. Toosi University of Technology Tehran, Iran, email: vmohammadzadeh@email.kntu.ac.ir

**Mahdi Aliyari Shoorehdeli**

Department of Electrical Engineering K. N. Toosi University of Technology Tehran, Iran, email: aliyari@kntu.ac.ir

**Abstract**

False and nuisance alarms in industrial fault detection systems are often triggered by uncertainty, causing normal process variable fluctuations to be erroneously identified as faults. This paper introduces a novel encoder-based residual design that effectively decouples the stochastic and deterministic components of process variables without imposing detection delay. The proposed model employs two distinct encoders to factorize the latent space into two orthogonal spaces: one for the deterministic part and the other for the stochastic part. To ensure the identifiability of the desired spaces, constraints are applied during training. The deterministic space is constrained to be smooth to guarantee determinism, while the stochastic space is required to resemble standard Gaussian noise. Additionally, a decorrelation term enforces the independence of the learned representations. The efficacy of this approach is demonstrated through numerical examples and its application to the Tennessee Eastman process, highlighting its potential for robust fault detection. By focusing decision logic solely on deterministic factors, the proposed model significantly enhances prediction quality while achieving nearly zero false alarms and missed detections, paving the way for improved operational safety and integrity in industrial environments.

1- **Introduction**

In industrial systems, the top priority is ensuring the plant operates smoothly. For this purpose, modern facilities are equipped with numerous sensors strategically placed to monitor various physical and environmental conditions. These sensors communicate through specific channels to provide real-time monitoring data. When a process variable crosses a predefined threshold, an alarm is activated to signal an abnormal condition. These alarms alert operators to any issues, indicating unusual plant behavior. Timely detection and response to faults are essential for maintaining cost efficiency, ensuring the safety of personnel and the facility, and upholding product quality. Delays in addressing these faults can lead to severe consequences, including potential injuries or fatalities [1].

The design process for fault detection systems typically involves two crucial steps: residual design and residual analysis. The residual is defined as a signal that must be activated solely by abnormal conditions, despite industrial systems being influenced by various factors other than faults. There are numerous methods for residual design as discussed in [2-5]. These techniques can generally be divided into two main categories: model-based and signal-based, or rather data-driven, approaches. Model-based residual design often faces practical challenges, as for many real-world systems, either obtaining accurate mathematical models is difficult or modeling is time-consuming and their validity is not always guaranteed; any changes in the process necessitate revisions to the model [2]. The complementary step involves analyzing the designed residual. Residual analysis methods search for patterns in the residual to differentiate between various system states. The success of this step is closely tied to the quality of the residual. Analysis techniques include statistical methods and pattern recognition methodologies [6, 7], but the simple limit check remains the most commonly used method in industrial settings [2]. While this method is advantageous due to its interpretability and ease of implementation, it also introduces certain challenges.

The most important challenges with this method is the proper selection of the threshold which directly affects the false and missed alarm rates. A false alarm occurs when an alarm is triggered despite no abnormality being present in the process, while a missed alarm happens when no alarm is triggered despite a fault occurring. In the simple limit checking method, if the threshold is set incorrectly, normal fluctuations of measured variables may lead to a high number of false alarms during fault-free operation or significantly increase missed alarms when a fault is present. This issue is primarily due to the uncertainty in process variables and the design of the threshold [8]. To address this within the framework of fault detection system design, there are two main approaches. One approach involves modifying the decision logic for raising alarms by introducing changes such as implementing a waiting period or incorporating a dead zone for process variables to either trigger or clear alarms [9-15]. Another approach is to monitor a function of the process variables, known as filtered variables, instead of the variables themselves [16-19]. The main objective of this function is to derive a meaningful representation of the data, providing enriched information about the system's health.

Given the simplicity of limit checking methods for fault detection, it is beneficial to focus on improving the residual signal design. From this perspective on fault detection system design, the success of monitoring systems depends significantly on the design methods. By revisiting the residual design process, the residual can be seen as a representation of the process variables, encapsulating sufficient information about the system's health. Since this representation is not predetermined, it can be referred to as a latent representation. This approach aligns closely with representation learning in Machine Learning (ML), specifically Deep Learning (DL) domain, where the goal is to find an intermediate representation that facilitates decision making [6, 7, and 20]. Given their success in applications like natural language processing (NLP), computer vision (CV) and time series analysis, they are promising tools for enhancing the design of FDI systems. It is worth mentioning that the integration of ML algorithms into fault detection systems is not limited to recent years, and it has been developing for decades under the intelligent fault detection (IFD) title. However, they are initially used as an agent which make a decision. For instance, signal processing algorithms are employed to extract sensitive features from process variables, then the results are fed into an intelligent system like Artificial Neural Networks (ANN) [21-23], Support Vector Machine (SVM) [24, 25], and K-nearest Neighborhood (KNN) [26] for categorizing systems' health status. Although, the reported works proof their capability of detecting anomalies, it is in contrast with the core functionalities of ML algorithms in extracting meaningful information from the raw variables [20].

In [27] Convolutional Neural Networks (CNN) is used to extract information from time-frequency features. The authors introduce a novel network architecture, WPT-CNN, designed for end-to-end IFD. This network uniquely integrates wavelet packet transform (WPT) capabilities into the CNN framework, enabling simultaneous time-frequency analysis and fault classification without the need for manual signal processing. Simplified Shallow Information Fusion-Convolutional Neural Network (SSIF-CNN) is introduced in [28], where time domain and frequency domain features were extracted from both training and testing samples prior to being input into the SSIF-CNN model, then feature maps generated by the hidden layers were transformed into corresponding feature sequences using global convolution operations. The fault detection will be performed based on the concatenation of feature maps into 1-D vector. In [29] an adaptive deep convolution neural networks (ADCNN) is presented leveraging a two-dimensional visualization of raw acoustic emission (AE) signals to automate and enhance the feature extraction and classification processes. The model employs a discrete wavelet packet transform (WPT) to process the AE signals and introduce a new evaluation metric called the Degree of Defectiveness Ratio (DDR), resulting in a visualization referred to as a DDRgram. This approach captures valuable information distributed across different frequencies, enabling better insight into the health conditions of bearings. In [30] a diagnostic framework, called ASM1D-GAN, is introduced which address the problem of lacking supervised data. The ASM1D-GAN consists of a 1D convolutional neural network, Generative Adversarial Networks (GANs), and a fault classifier, thereby combining both data generation and diagnostic processes. In [31] a multilayer wavelet attention convolutional neural network (MWA-CNN) designed for noise-robust machinery fault diagnosis. This novel layer maps time-domain signals to wavelet space to extract valuable information using a learnable convolutional layer. The framework alternates between DWA-Layers and standard convolutional layers to perform signal decomposition and feature learning, effectively embedding a multi-resolution analysis algorithm within the CNN structure. In [32] a combination of short-time Fourier transform (STFT) with CNN is given. One dimensional vibration data is first transformed to time-frequency images using STFT, which helps to better capture the relevant features for diagnostic purposes. These time-frequency images are then inputted into the STFT-CNN model for fault feature learning and fault identification.

Another direction in the literature is the use of auto-encoders (AE). Auto-encoders consist of an encoder layer and a decoder layer, learning a deterministic mapping that compresses the original input space into a lower-dimensional representation and then reconstructs it back to the original form. In one of the earliest works, [33] proposes a three-layered stacked auto-encoder to extract fault-related information from the spectrum of input variables. Similarly, an integrated deep fault recognizer model based on stacked de-noising auto-encoders (SDAE) is presented in [34] that is used to both de-noise raw signal noise and extract fault features for diagnosing bearing rolling faults and gearbox faults, trained in a greedy layer-wise manner. In [35] two step training procedure is given that utilize a regularized form of SDAE for feature learning. It establishes a deep hierarchical structure using greedy training rules, with an emphasis on sparsity representation and data corruption to extract high-order characteristics and enhance robustness during learning. Considering fault severity classification, a method is given in [36] that is based on a deep sparse auto-encoder network trained with noise-augmented sample expansion, enabling unsupervised learning of the underlying structure and characteristics of the data. The proposed solution involves stacking multiple sparse auto-encoders with a classifier layer to create a deep sparse auto-encoder network that can intelligently identify fault severity. As an important variants of auto-encoders, a VAE-based approach is given in [37] which solve the issues with single feature thresholding for non-stationary nature of vibration signal. They propose a VAE based representation learning which extract explanatory factors behind normal working condition, and consider the reconstruction error of the decoder network as residual signal. A similar work is introduced in [38] exploiting VAE-based architecture to monitor the performance of ball screw component. First, VAE model is trained on collected data from normal operating condition, then the reconstruction error of trained VAE is regarded as residual signal. Subsequently, kernel density estimation is employed to assess the probability distribution of obtained residual signal within a dynamic sliding window. The evaluation of deterioration is then conducted by aggregating the probability distribution values that surpass a pre-defined threshold. In [39] a semi-supervised learning approach is given which uses VAE for learning a representation based on a vast amount of unlabeled data. The obtained representation is then fine-tuned on small set of labeled data. In [40] a novel method is introduced that is tailored for multivariate sensor data using graph-based neural networks. The proposed strategy uses graph auto-encoders to provide effective fault detection by identifying complex relationships and dependencies among sensors, facilitating anomaly detection to spot abnormal patterns that may indicate faults. It utilizes attention mechanisms to prioritize relevant sensor nodes, enabling the capture of their spatial and temporal dependencies for enhanced fault diagnosis.

In nearly all aforementioned works and others listed in [41] and [42], fault detection system is defined in two separate stage, one is learning the representation and second is the analysis of the learned representation. The analysis step is inevitable, since to verify the quality of the learned representation, an additional projection head should be added on top of the auto-encoder's bottleneck layer to classify system health statuses, bringing additional computational cost as compared to the thresholding. What is more, there is no guarantee that the model captures health-state-related information at the representation level through unsupervised learning. This will be provided during the fine-tuning stage where the labeled data is incorporated to adjust the learned parameters to classify systems health status. This is because auto-encoders are unsupervised and aim to encode one space into another, whereas it is crucial from a system perspective to identify the primary factors behind data variations. This is particularly important in industrial alarm systems, where normal fluctuations might be misinterpreted as faults. There are some works which pay attention to the nature of representation the model learns. In [43] a VAE-based architecture is adopted to foster a compact and informative latent representation, enhancing the detection and segmentation of previously unseen fault types. The model uses supervision data to shape the latent representation to be ensured that the model can accurately distinguish between normal and abnormal condition. Another work, [44] exploiting the capacity of multi-modal learning to fuse vibration and acoustic signal at the representation level for spark plug fault detection. Given labeled data, it uses an auto-encoder based architecture to maximally align the representation of both modalities belonging to the same health state. To address the stated challenges, a novel encoder-based architecture is used to learn a factorized latent space that can distinguish between deterministic and stochastic factors of process variables while keep the analyzing method as thresholding. Separating the stochastic and deterministic representation of data can significantly enhance the prediction quality, as the random fluctuation of signals is removed and the decision logic is only performed based on the deterministic factor. Our encoder-based model features two separate encoders: one for deterministic and one for stochastic representations of variables. To ensure identifiability, we incorporated the following constraints into the objective function: a smoothness constraint for the deterministic encoder to ensure consistency, minimization of the KL divergence between the stochastic representation and a standard Gaussian, and an orthogonal loss to maintain the independence of the information captured by each encoder. The decision logic for fault detection is only performed on deterministic representation using comparison with a pre-defined threshold.

The illustration of the proposed method is depicted in Figure 1.

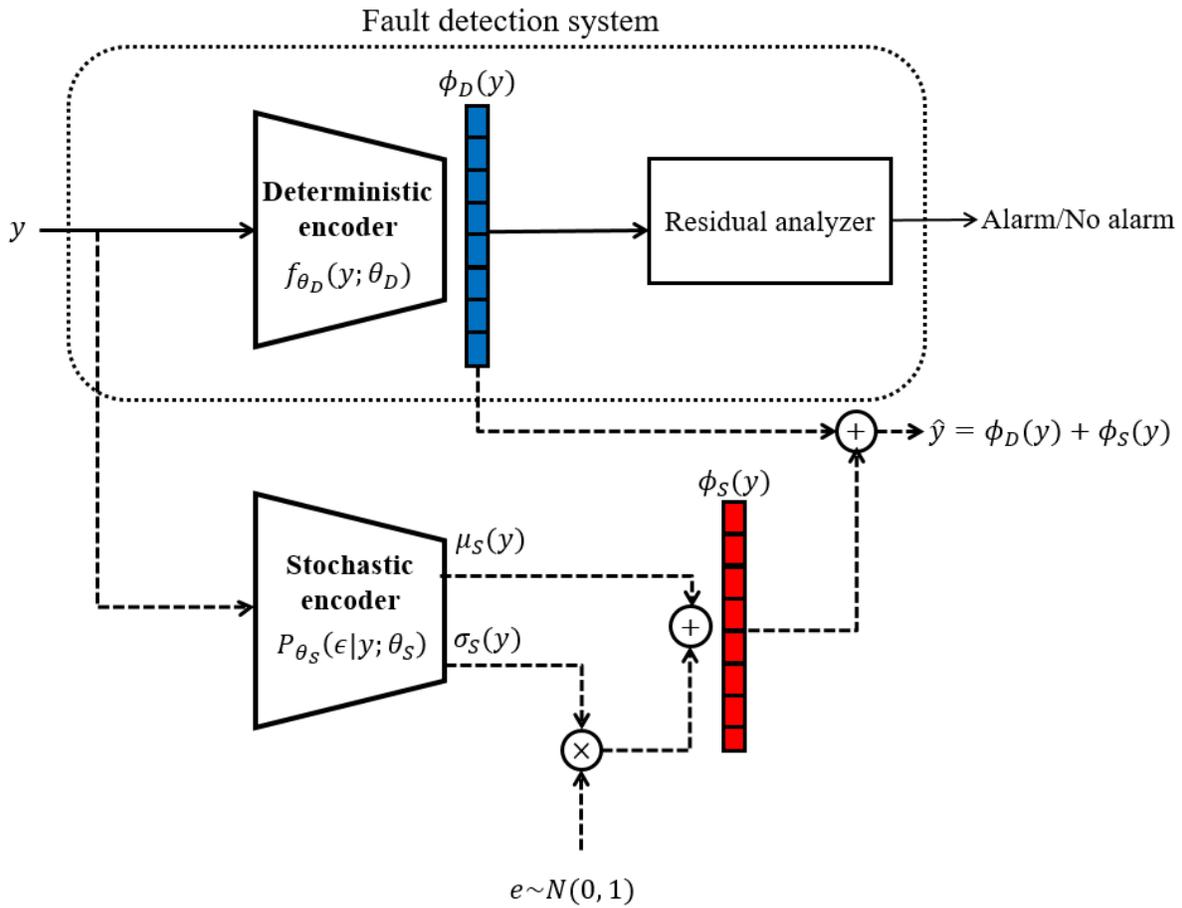

Figure 1 – The schema of proposed approach. The dashed arrow lines are removed at inference stage and dot box is added.

The proposed schema, illustrated in Figure 1 without the dotted box, is trained using normal data. During training, the model learns to extract the deterministic component and inherent variations within the data. Once training is complete, the path to the stochastic encoder, represented by dashed arrow lines, will be removed, and the dotted box will be activated to make decisions based on the deterministic representation of the input signal as a residual signal.

2- **Problem definition**

It is assumed that the measurement $y$ can be decomposed into deterministic part $x$ and stochastic part $\epsilon$, i.e. ($y = x + \epsilon$), and the goal is to design a model to capture it. For this purpose, we present an encoder-only architecture designed to extract both deterministic and stochastic representations from process variables. The model employs two distinct encoders, each tailored to address different aspects of data representation. The deterministic encoder focuses on capturing stable, consistent representations, enhanced by the application of smoothness constraints. In parallel, the stochastic encoder models the probabilistic aspects of the data, with KL divergence minimization ensuring alignment with a standard Gaussian distribution. To maintain the independence of the information captured by each encoder, we incorporate an orthogonal loss. The following subsections detail the structure and functionality of each component, emphasizing the innovative aspects that distinguish our approach from traditional methods.

## 2-1- Deterministic encoder

The deterministic encoder in our model is designed to capture stable and consistent representations of the input data. This encoder operates as a traditional encoder within an auto-encoder framework, albeit with specific modifications to ensure the deterministic nature of the process variables. An auto-encoder consists of two main parts: an encoder and a decoder. The encoder compresses the input data into a lower-dimensional representation, while the decoder attempts to reconstruct the original data from this compressed form. Omitting the decoder model, the encoder model can be implemented using a simple MLP network, depicted in Figure 2.

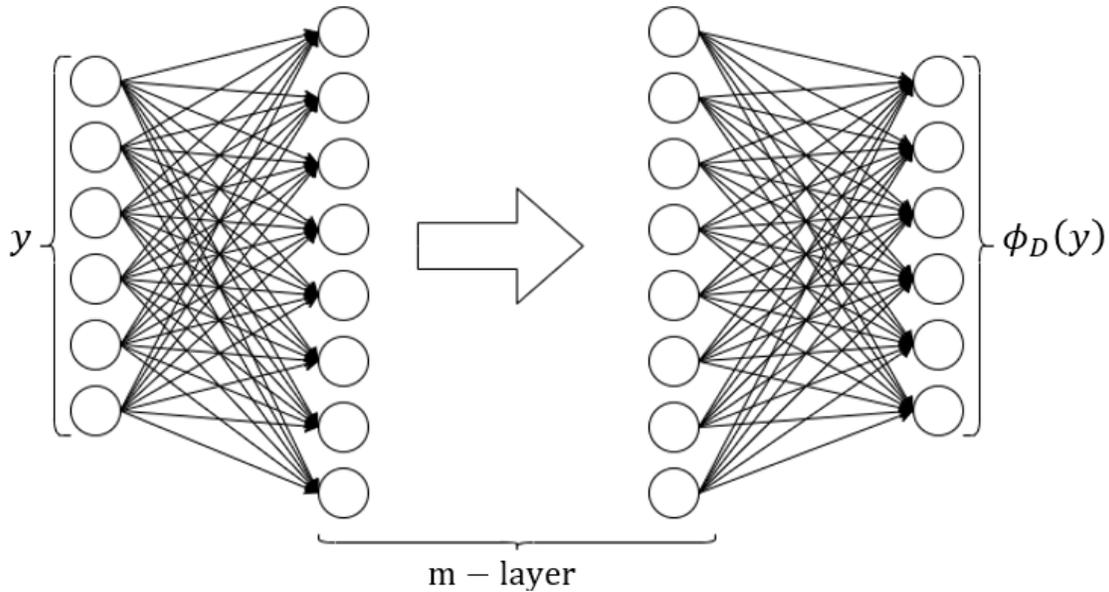

Figure 2 – Deterministic encoder capturing the deterministic representation from the measurement $y$

Starting from the input layer, the mathematics behind an m-layered MLP network would be as following:

$$h^{(1)} = f^1(W^{(1)}y + b^1), h^{(k)} = f^{(k)}(W^{(k)}h^{k-1} + b^k), k = 2, \ldots, m \tag{1}$$

The deterministic representation is computed as the output of network:

$$\phi_D(y) = W^{m+1}h^m + b^m \tag{2}$$

### 2-1-1- Smoothness implies determinism

As previously discussed, the representation $\phi_D(y)$ is straightforwardly computed using the specified relationships. However, a significant point of contention lies in the nature of this representation. Although we assume that the measurement $y$ can be decomposed into a deterministic part $x$ and a stochastic part $\epsilon$, a pertinent question arises: how can we ensure that $\phi_D(y)$ is truly deterministic?

Determinism refers to the property where the output of a system is entirely determined by its input, without any randomness, indicating predictability. To achieve determinism in a system, the system's behavior must be smooth. Smoothness in this context means that the system's functions are continuous and differentiable, without abrupt changes or discontinuities. This property is essential for the precise prediction of future states based on current conditions. Research has shown that smoothness in the data or system behavior indicates determinism because it allows the use of differential equations to describe the system's evolution. These equations ensure that small changes in conditions result in predictable changes in outcomes, aligning with the principles of deterministic systems [45]. So, by definition, to ensure that $\phi_D(y)$ is deterministic, we must demonstrate that small changes in $y$ result in small, consistent changes in $\phi_D(y)$. The equivalent probabilistic perspective is that the likelihood of the deterministic representation given the measurements should be stable under small perturbations in $y$.

$$p(\phi_D(y+\delta)|y) \approx p(\phi_D(y)|y) \tag{3}$$

***Theorem-1***: *If a deterministic representation $\phi_D(y)$ is smooth, then the likelihood function $P(y|\phi_D(y))$ will be robust to small perturbations in the input $y$. This implies that adding a smoothness loss term to the cost function ensures that $\phi_D(y)$ is deterministic.*

**Proof**: Given the measurement model $y = x + \epsilon$, where $y$ is measurement, $x$ denotes deterministic part and $\epsilon$ is modeled as random noise. The deterministic representation $x = \phi_D(y)$ should map $y$ into $x$. A function $\phi_D(y)$ is smooth if small perturbation in $y$ leads to small change in $\phi_D(y)$. Formally, this can be expressed using Lipchitz conditions:

$$|\phi_D(y+\delta) - \phi_D(y)| \leq L|\delta| \tag{4}$$

Where $L$ is the Lipschitz constant. To formalize it, we can introduce a Gaussian prior on the changes in the deterministic representation:

$$\phi_D(y+\delta) - \phi_D(y) \sim N(0, \sigma^2 I) \tag{5}$$

Where $\sigma^2$ is small variance. Incorporating it into likelihood function will result in:

$$P(y|\phi_D(y)) \propto \exp\left(-\frac{1}{2\sigma^2}\left(\phi_D(y+\delta) - \phi_D(y)\right)^2\right) \tag{6}$$

The negative-log-likelihood introduce the smoothness loss term:

$$L_{smoothness} = |\phi_D(y+\delta) - \phi_D(y)|^2 \tag{7}$$

## 2-2- Stochastic encoder

The stochastic encoder is designed to model the variable and probabilistic aspects of the input data. Unlike the deterministic encoder, which captures stable and consistent features, the stochastic encoder accounts for uncertainty and inherent variability in the data by encoding it as a probability distribution. The stochastic encoder serves to represent the uncertainty in the data through a distribution rather than a single deterministic point. This is particularly useful for tasks requiring probabilistic reasoning or when modeling complex phenomena with inherent randomness. Let $\phi_S(y)$ represent the stochastic encoder's output for an input $y$. Typically, the encoder maps the input to a latent variable that follows a probability distribution, often assumed to be Gaussian for mathematical convenience and ease of optimization. Keeping this in mind, the stochastic encoder produces the parameters for this distribution. The mathematical relationships for the stochastic encoder are similar to those of the deterministic encoder, except it includes two projection heads from the latent space to the Gaussian distribution parameters: one for the mean and another for the logarithm of the standard deviation. Figure 3 provides an illustration of the stochastic encoder

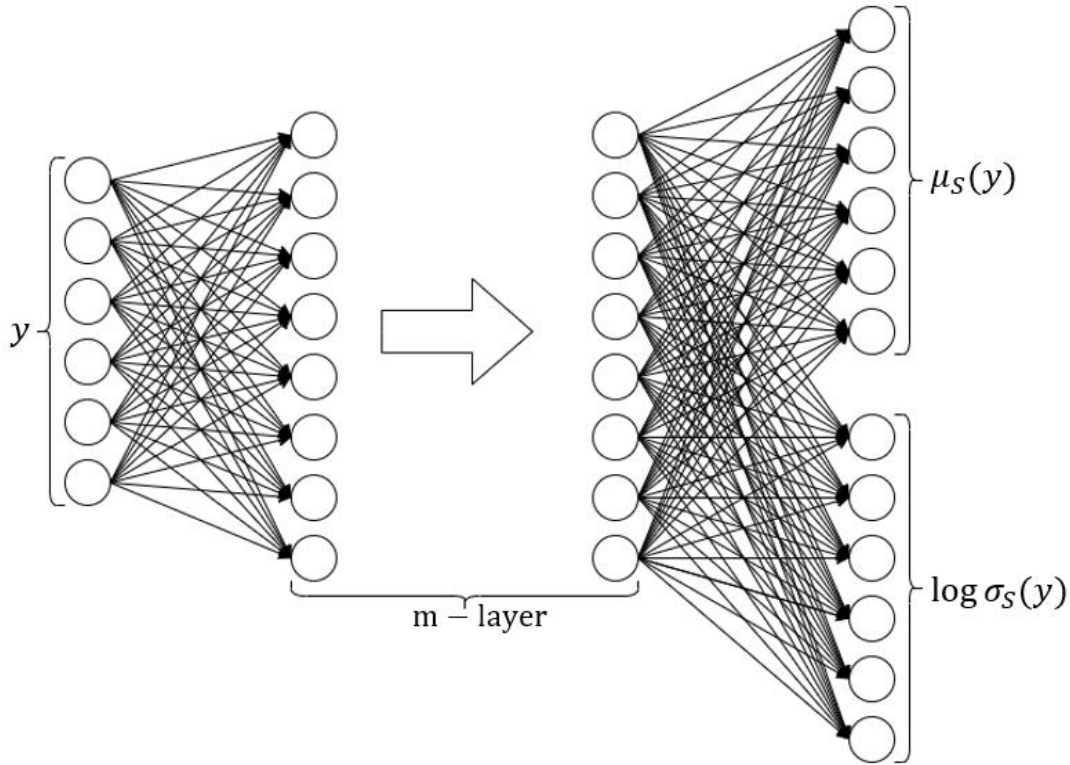

Figure 3 – Stochastic encoder capturing the inherent variations behind data from the measurement $y$

The stochastic representation $\phi_S(y)$ is simply obtained through parametrization tricks:

$$e \sim N(0, I) \rightarrow \phi_S(y) = \mu_S(y) + e\sigma_S(y) \tag{8}$$

## 3-3- Objective function

The associated probabilistic graphical model with proposed model is depicted in Figure 4. The joint distribution over the involved variable is defined as follows:

$$p_\theta(y, x, \epsilon) = p_\theta(y|x, \epsilon)p_\theta(x)p_\theta(\epsilon) \qquad (9)$$

Where $P_\theta(\epsilon)$ follows a standard Gaussian distribution ($N(0, I)$), and $P_\theta(x)$ is the prior distribution defined in relationship (2).

The measurement model is also simply defined as the sum of the deterministic and stochastic representations:

$$\hat{y} = \hat{x} + \hat{\epsilon} = \phi_D(y) + \phi_S(y) \qquad (10)$$

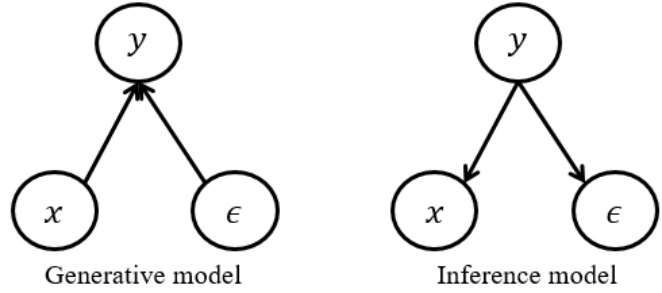

Figure 4 – The probabilistic graphical model corresponds to proposed model

Thus the conditional distribution $P_\theta(y|x, \epsilon)$ is parametrized as follows:

$$p_\theta(y|\phi_D(y), \phi_S(y)) \sim N(\phi_D(y) + \mu_S(y), \sigma_S(y)) \qquad (11)$$

The corresponding likelihood function is formulated as:

$$\log p_\theta(y|\theta) = \log \int p_\theta(y, x, \epsilon|\theta) dx d\epsilon = \log \int p_\theta(y|x, \epsilon, \theta) p_\theta(x) p_\theta(\epsilon) dx d\epsilon$$

$$\geq E_q\left[\log \frac{p_\theta(y|x, \epsilon, \theta) p_\theta(x) p_\theta(\epsilon)}{q_\lambda(\epsilon|y) q_\phi(x|y)}\right] = L_\theta$$

Where $q_\lambda(\epsilon|y) q_\phi(x|y)$ is associated with inference model for obtaining the deterministic and stochastic representation. Since $x$ is considered as deterministic part, the conditional distribution $q_\phi(x|y)$ is a degenerative distribution defined as:

$$q_\phi(x|y) = \delta(x - \phi_D(y)) \qquad (12)$$

The conditional distribution $q_\lambda(\epsilon|y)$, as custom in stochastic encoders, is parametrized using a Gaussian distribution whose mean and covariance are governed by $\mu_S$ and $\sigma_S$. So, we will have:

$$L_\theta = E_{q_\lambda}\left[\log \frac{p_\theta(y|\phi_D(y), \epsilon, \theta) p_\theta(\phi_D(y)) p_\theta(\epsilon)}{q_\lambda(\epsilon|y)}\right]$$

$$= E_{q_\lambda}[\log p_\theta(y|\phi_D(y), \epsilon, \theta)] + E_{q_\lambda}[\log p_\theta(\phi_D(y))] - D_{KL}(N(\mu_S, \sigma_S)|N(0, I))$$

$$L_\theta = -E_{q_\lambda}\left[\sum_{k=1}^{N} \log p_\theta(y_k|\phi_D(y_k), \epsilon_k, \theta)\right] + \sum_{k=1}^{N} |\phi_D(y_k) - \phi_D(y_{k-1})|^2$$

$$+ D_{KL}(N(\mu_S, \sigma_S)|N(0, I))$$

Importantly, to ensure that there is no shared information between $\phi_D(y)$ and $\phi_S(y)$, they must be orthogonal. This can be achieved by incorporating a similarity measure $K(\phi_D(y), \phi_S(y))$. In this work, we use dot product. The complete cost function for parameter optimization is as the follows:

$$L_\theta = \phi_D(y)^T \phi_S(y) - E_{q_\lambda}\left[\sum_{k=1}^{N} \log p_\theta(y_k | \phi_D(y_k), \epsilon_k, \theta)\right] \\ + \lambda_{smoothness} \sum_{k=1}^{N} |\phi_D(y_k) - \phi_D(y_{k-1})|^2 + D_{KL}(N(\mu_S, \sigma_S) | N(0, I)) \quad (13)$$

Both the deterministic and stochastic encoders are configured so that the resulting objective function is minimized.

3- **Simulation**

In this section, we present the simulation framework utilized to evaluate the efficacy of the proposed fault detection model. The evaluation is conducted through two distinct examples: a numerical example designed to test fundamental aspects of the model and an industrial benchmark, Tennessee Eastman Process (TEP).

The numerical example allows for controlled experimentation, enabling us to systematically investigate specific features of the proposed model, including its ability to detect anomalies and diagnose faults under varying conditions. By utilizing simplified scenarios, we can assess the performance of the model while manipulating parameters to observe its responsiveness and robustness. In contrast, the Tennessee Eastman Process serves as a complex, real-world benchmark that reflects the intricacies and challenges faced in industrial applications. As a well-established testbed in the field of process systems engineering, the TEP provides a rich dataset that includes various operational states and fault scenarios. This enables a comprehensive evaluation of the model's performance in a realistic setting, highlighting its practicality and applicability in real-world situations.

**3-1- Numerical simulation**

In this subsection, we detail the numerical simulation example devised to evaluate the performance of the proposed approach. This example is centered around the generation of a dataset that mimics normal operating conditions, which is critical for establishing a baseline for anomaly detection. For this simulation, we generated 10,000 measurement samples from a multivariate Gaussian distribution with a dimensionality of 16, characterized by the parameters $N(2, 1)$ for each dimension. In this context, every variable within the multivariate distribution has a mean of 2 and a standard deviation of 1, reflecting the expected normal functioning state of the system across its 16 dimensions.

To optimize the performance of the proposed model, the hyper-parameters of the neural network were carefully determined through a grid search methodology. This systematic approach involves testing a diverse range of hyper-parameter configurations, enabling the identification of the most effective settings for the network architecture. The adopted architecture takes a nonlinear mapping of the input variables as the input of deterministic and stochastic encoder. The shape of nonlinear mapping, deterministic encoder, and stochastic encoder are (16, 100, 50), (50, 85, 16), and (50, 65, 16) respectively. To optimize the performance of the proposed neural network model, we utilized the Adam optimizer, and the learning rate was set to 0.001. The learning curve of the trained model is depicted in Figure 5. The range of obtained value indicate that the incorporated terms into the cost function (13) is suitably approach to zero.

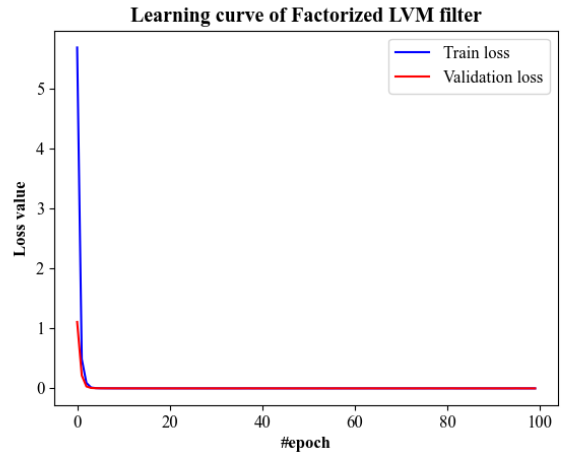

Figure 5 – The value of training loss and validation loss during per epochs

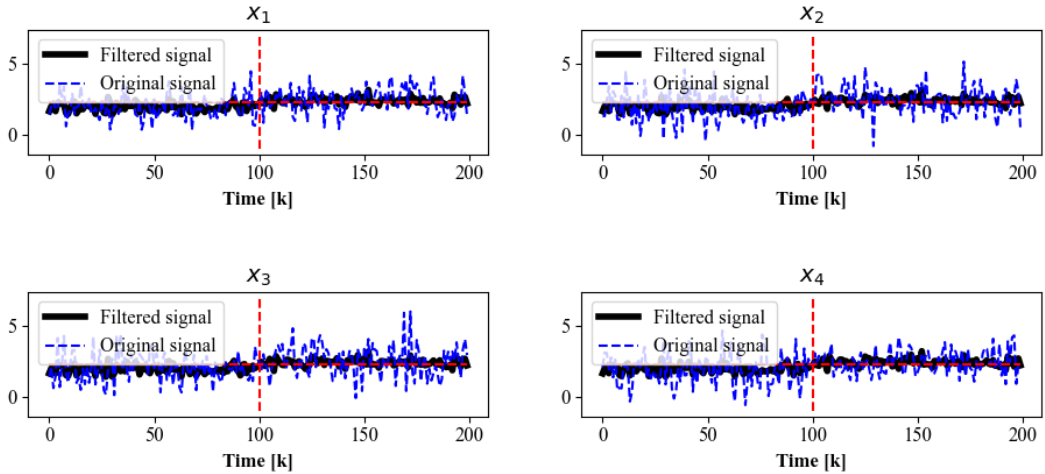

Figure 6 – The output of proposed model for Gaussian fault with mean 2.5 and standard deviation 1

Now, to examine the diagnostic capabilities of the proposed model, we consider that the fault follows the Gaussian distribution with the same covariance matrix but different mean vectors. The mean vector values of fault state are selected relative to the normal operating values, small deviation (0.5 unit), medium deviation (1 unit) and large deviation (2 unit). The optimal threshold for the given condition is calculated as 2.25, 2.5, and 3. In all scenarios, the trained model is tested by the concatenation of normal and abnormal state along the time axes, and it is assumed that the fault is occurred at time step 100. First, we consider the case at which the difference between the mean of normal and abnormal data is small. Given the closeness of statistics, the detection performance using a simple threshold is hard. The time-based original and filtered signals are given for some dimensions in Figure 6. The normal and abnormal state is separated by the vertical dashed line shown in the picture, at time instance 100, where the left side is normal data and the other is abnormal state. The horizontal dashed is also the optimal threshold. To make the performance of the proposed model clear, consider the distribution of data before and after filtering as Figure 7:

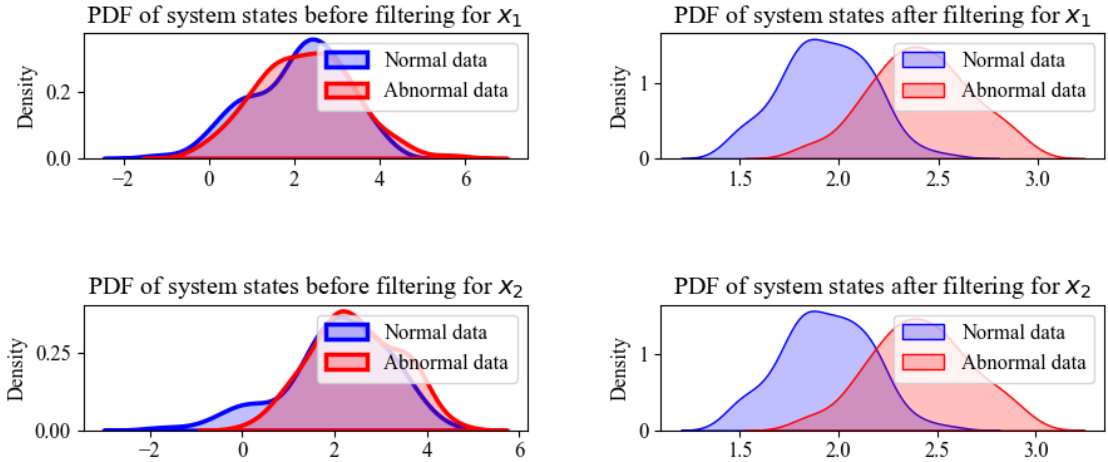

Figure 7 – The distribution of normal and abnormal state in original and filtered form in the presence of Gaussian fault with mean 2.5 and standard deviation 1

Based on the given optimal threshold 2.25, FAR and MAR for the original signal are computed as 0.46 and 0.44 respectively, while for the filtered one those are computed as 0.07 and 0.1 respectively, indicating the effectiveness of the proposed method. Subsequently, for Gaussian fault data with mean 3 and standard deviation 1, Figure 8 and 9 are reported.

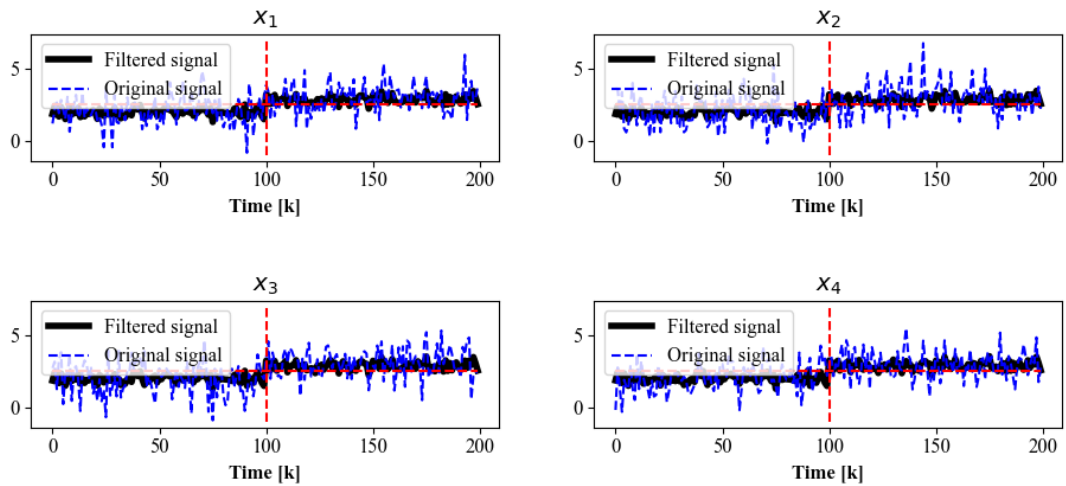

Figure 8 – The output of proposed model for Gaussian fault with mean 3 and standard deviation 1

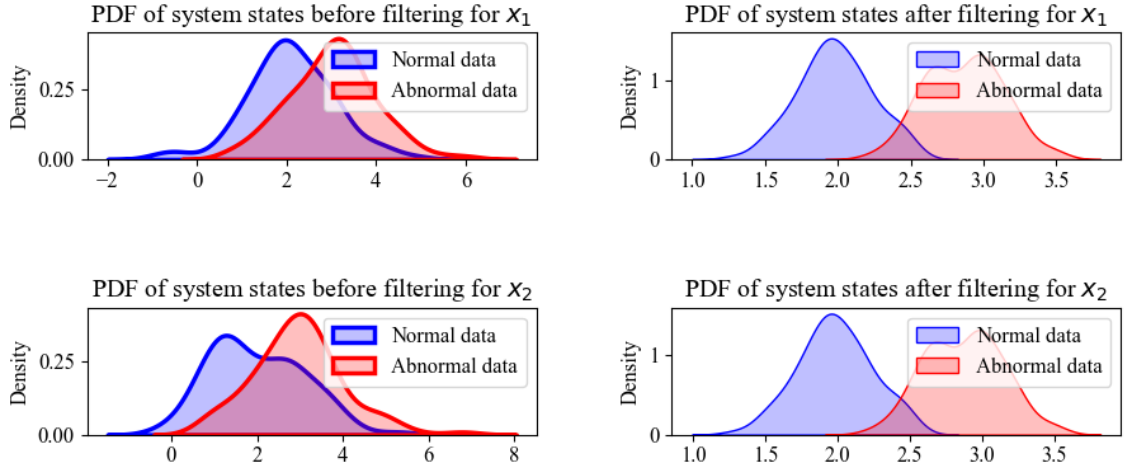

Figure 9 – The distribution of normal and abnormal state in original and filtered form in the presence of Gaussian fault with mean 3 and standard deviation 1

Using the optimal threshold of 2.5, the False Alarm Rate (FAR) and Missed Alarm Rate (MAR) for the original signal are calculated to be 0.2 and 0.25, respectively. In contrast, for the filtered signal, these rates are found to be 0.01 and 0.04, respectively. Following this, Figures 10 and 11 present the results for Gaussian fault data with a mean of 4 and a standard deviation of 1.

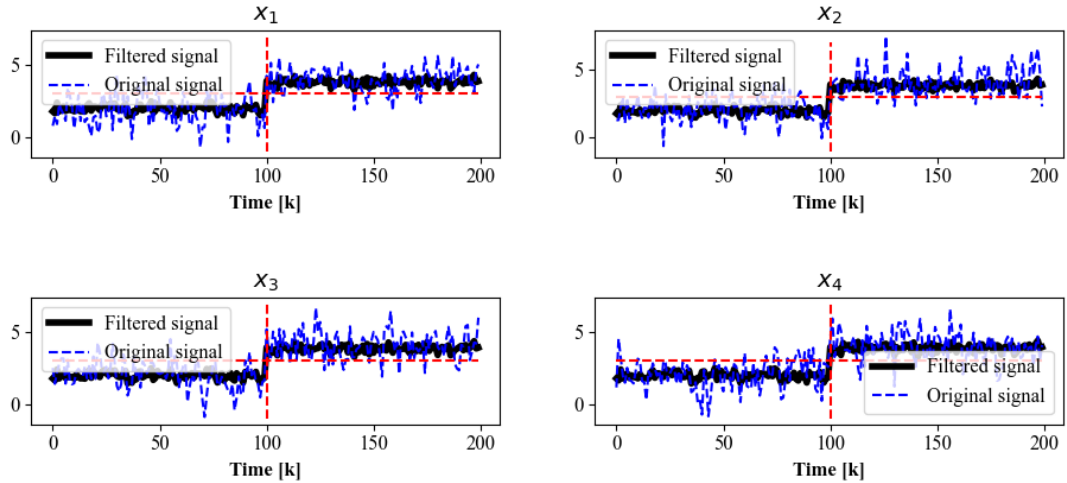

Figure 10 – The output of proposed model for Gaussian fault with mean 4 and standard deviation 1

With an optimal threshold of 3, the False Alarm Rate (FAR) and Missed Alarm Rate (MAR) for the original signal are determined to be 0.1 and 0.11, respectively. In comparison, both rates for the filtered signal are measured at 0.

Simulation results clearly show that the proposed method is capable of capturing the variation behind the data and remove fluctuations from the signal, leading to remarkable reduction in false and alarm rates, and increasing the reliability of IFD system. To ensure the performance of the designed system, an analysis of detection delay should be also performed, since it is likely this performance is obtained in exchange for increasing the computational efforts of encoder model. For this purpose, we measure the CPU inference time required for the computation of residual and comparing it with a predefined threshold to raise an alarm. We consider the case where fault state follows Gaussian distribution with mean 4 and standard deviation 1 in all dimension. The results are depicted in Figure 12. The inference time, as the detection delay, is virtually 0.0003 sec in average, which is suitable for a wide range of systems.

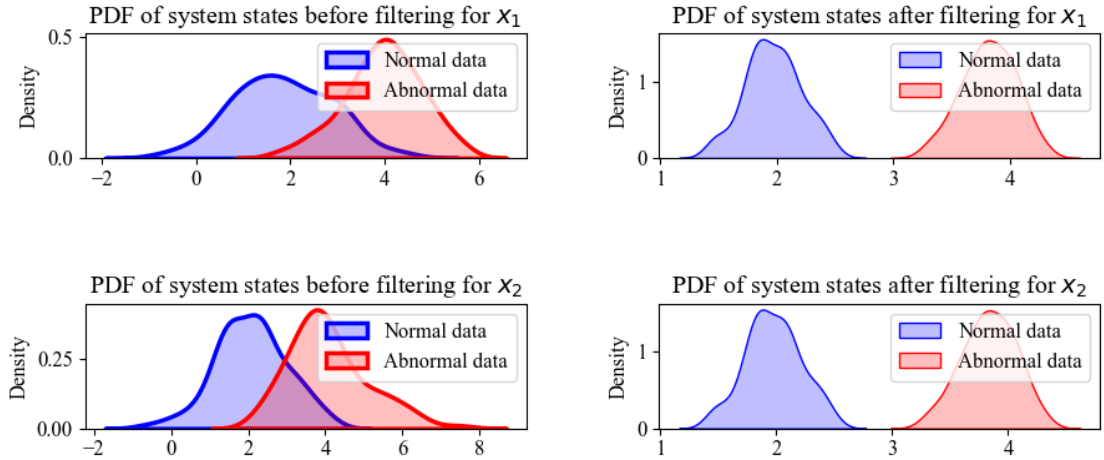

Figure 11 – The distribution of normal and abnormal state in original and filtered form in the presence of Gaussian fault with mean 4 and standard deviation 1

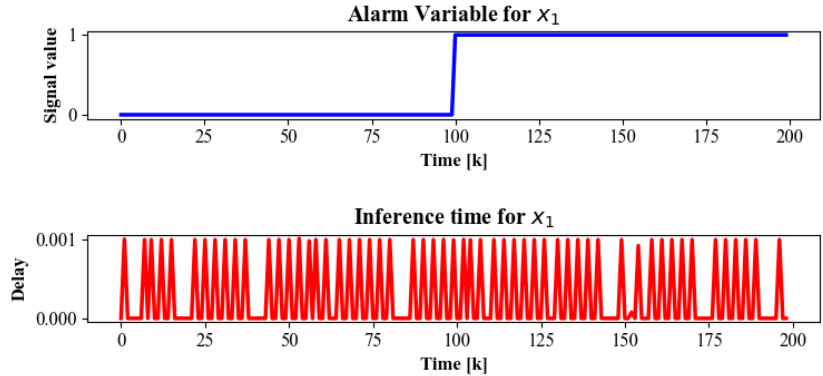

Figure 12 – The alarm variables and computational time for raising alarms

The obtained results are summarized in Table 1, where it is assumed that the normal operating condition follows a Gaussian distribution with mean 2 and standard deviation 1 in all dimensions.

Table1 – The obtained result for numerical example

| State | Original signal | | | Filtered signal (proposed model) | | |
|---|---|---|---|---|---|---|
| | FAR | MAR | Inference time (average) | FAR | MAR | Inference time (average) |
| $F_1 \sim N(2.5, 1)$ | 0.46 | 0.44 | 0 | 0.07 | 0.1 | 0.00032 |
| $F_2 \sim N(3, 1)$ | 0.25 | 0.2 | 0 | 0.01 | 0.04 | 0.00035 |
| $F_3 \sim N(4, 1)$ | 0.1 | 0.11 | 0 | 0 | 0 | 0.00030 |

### 3-2- Tennessee Eastman Process (TEP)

The Tennessee Eastman Process (TEP) introduced in [46] holding considerable importance in addressing process monitoring and fault diagnosis. The system's main components consist of a two-phase reactor, a condenser, a liquid/vapor separator, a recursive current compressor, and a stripper tower. In this system, gaseous reactants A, C, D, E, and an inert substance B are introduced into the reactor, leading to the formation of liquid products G and H, with material F being produced as a byproduct [47]. The flow diagram illustrating this process is shown in Figure 13.

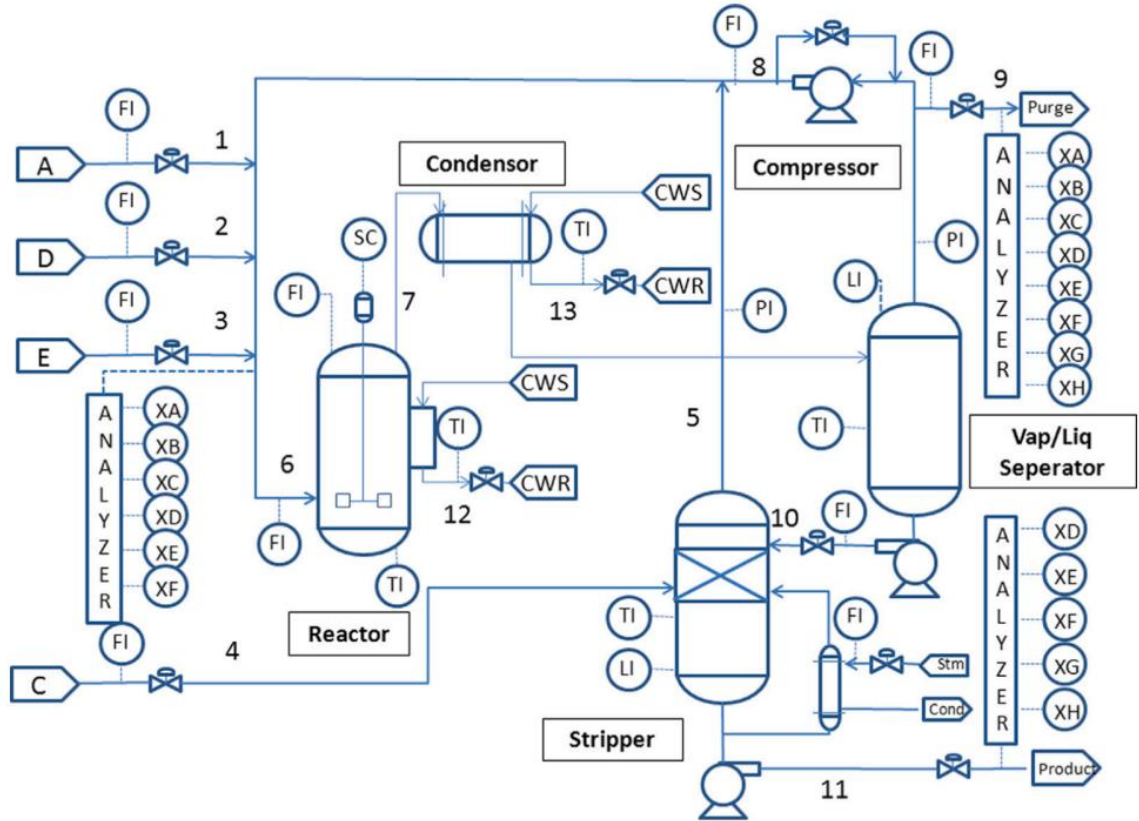

Figure 13 – Data flow of TE process [48]

In this study, we utilize simulation datasets generated in [47] to evaluate our proposed method for detecting faults. Although the dataset contains 52 process variables, we focus exclusively on measurement variables 1 to 22 for our analysis. To enhance the rigor of our evaluation, we introduce Gaussian white noise into the signals, which adds realism to the data. This dataset includes 1,800 samples for both normal operating conditions and three different types of faults introduced at various points within the system: specifically, faults 2, 6, and 12. A complete listing of these variables along with their associated faults can be found in [47].

We train our proposed architecture using data collected from systems that operate normally, and we assess its effectiveness in detecting the three types of faults. The behaviors of these faults are generally similar for certain sensor variables, while the extent of change observed in others can vary. For instance, sensor variables 2, 3, 4, 5, 6, 12, 14, 15, and 17 show minimal impact from any faults, as illustrated by the original signal depicted in blue in Figures 14-16. In contrast, sensor 1, which does not respond to faults 2 and 12, clearly differentiates itself when fault 6 is present.

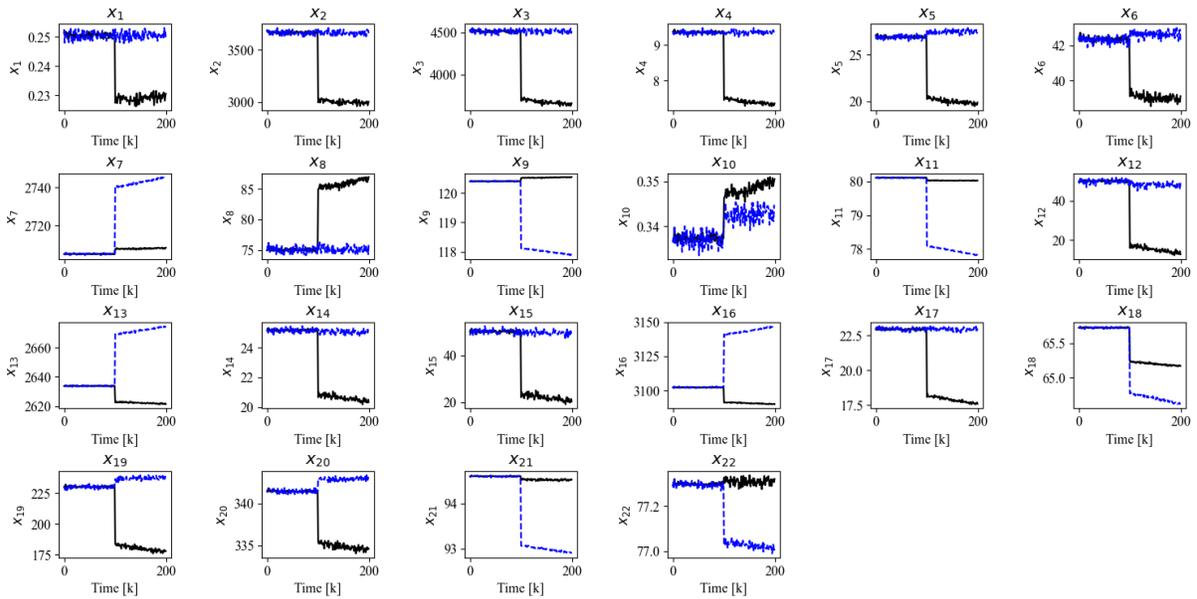

Figure 14 – The original and filtered signals of TEP variables for fault 2, where the blue line is the original signal and the black one is filtered signal (the output of deterministic encoder)

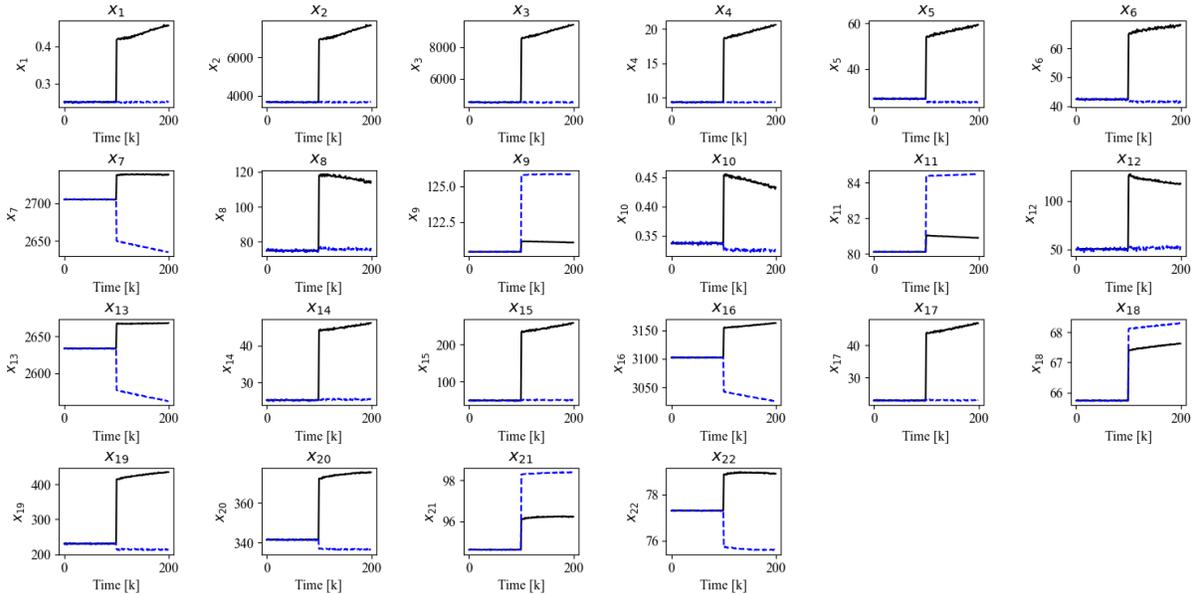

Figure 15 – The original and filtered signals of TEP variables for fault 12, where the blue line is the original signal and the black one is filtered signal (the output of deterministic encoder)

To illustrate the effectiveness of our proposed model, we concatenate the normal and abnormal states over time, under the assumption that faults occur at the time instance designated as 100. As depicted in Figures 14-16, our method successfully detects faults across all sensor variables. Specifically, for the signals that remain unaffected by the faults, the filtered signal (highlighted in black) shows a significant change in trend precisely at the time the faults manifest. This indicates that a straightforward threshold can facilitate effective decision-making, avoiding the necessity for complex decision logic. It is clear that by applying an appropriate threshold, we can achieve nearly zero false alarms and missed detections for nearly all variables, highlighting the effectiveness and reliability of the proposed method. This strong performance underscores the potential of our approach for effectively addressing fault detection in complex systems and sets the stage for future advancements in the field.

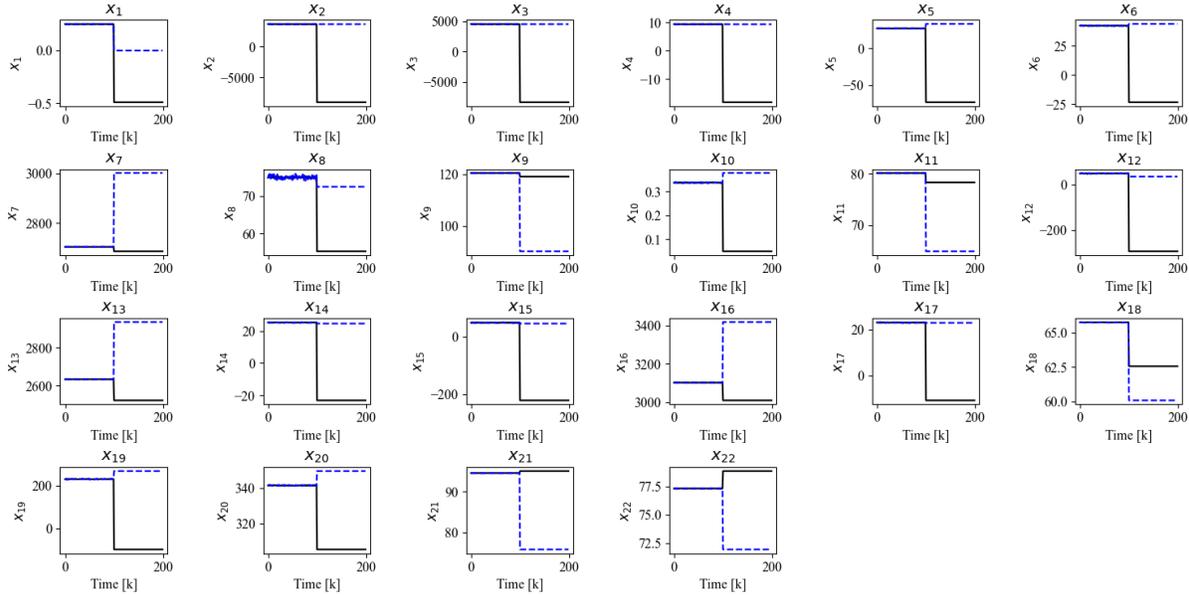

Figure 16 – The original (blue line) and filtered (black line) signals of TEP variables for fault 6, with the filtered signal representing the output from the deterministic encoder.

4- **Conclusion**

This work presents a novel encoder-based architecture specifically designed to improve fault detection systems by addressing the inherent challenges associated with traditional methodologies. Many existing approaches treat fault detection in two stages: the initial learning of representations followed by an analysis of these representations. While this process is widely accepted, it introduces additional computational costs, particularly due to the need for an extra projection head that may or may not effectively capture health-state-related information. Our proposed architecture circumvents these limitations by utilizing a factorized latent space that distinctly separates the deterministic and stochastic components of process variables. This is accomplished using two separate encoders that focus on learning and representing each component individually. The design ensures that the analysis can be performed directly on the deterministic representation, streamlining the process through simple threshold comparisons rather than complex classification layers.

To enhance the reliability and effectiveness of the model, several key constraints are incorporated into the training process. The deterministic encoder is subjected to a smoothness constraint to maintain consistency, while the stochastic representation is optimized to closely resemble standard Gaussian noise through the minimization of Kullback-Leibler divergence. Additionally, an orthogonal loss is employed to ensure that the information captured by each encoder remains independent, thereby increasing the clarity and quality of the predictions. By effectively filtering out random fluctuations and focusing solely on the deterministic factors that drive process variations, our model significantly enhances prediction quality in fault detection systems. This innovative approach not only simplifies the detection logic but also allows for more robust and efficient classification of system health states. The application of this encoder-based architecture, with its emphasis on clear representation learning and direct threshold-based analysis, shows great promise for advancing fault detection methodologies in industrial settings, ultimately aiming to improve operational safety and reliability.

1- **Reference**